\documentclass[conference,compsoc]{IEEEtran}
 \IEEEoverridecommandlockouts

\usepackage{amsmath,amssymb,amsfonts}
\usepackage{graphicx}
\graphicspath{{figures/}}
\usepackage{tikz}
\usetikzlibrary{positioning, arrows.meta, decorations.pathreplacing, fit, backgrounds}
\usepackage{subcaption}

\usepackage{booktabs}
\usepackage{tabularx}
\usepackage{array}
\usepackage{multirow}
\usepackage{todonotes}

\usepackage{textcomp}
\usepackage{xcolor}
\usepackage{xspace}
\usepackage[font=small,labelfont=bf]{caption}
\usepackage{url}
\usepackage{balance}
\usepackage{enumitem}

\usepackage{algorithm}
\usepackage{algpseudocode}

\usepackage{tcolorbox}

\usepackage[hidelinks]{hyperref}

\usepackage{cite}
\def\BibTeX{{\rm B\kern-.05em{\sc i\kern-.025em b}\kern-.08em
    T\kern-.1667em\lower.7ex\hbox{E}\kern-.125emX}}

\usepackage{fancybox}
\usepackage{xstring}
\makeatletter
\newcommand{\PP}[1]{%
  \vspace{2pt}%
  \noindent{\textbf{%
    \edef\@temp{\detokenize{#1}}%
    \IfEndWith{\@temp}{.}{#1}{#1.}%
  }}%
}
\makeatother

\newcommand{\summarybox}[1]{\vspace{0.5\baselineskip}\noindent\fbox{\parbox{0.94\columnwidth}{\small\textbf{Summary:}~#1}}}

\newcommand{\eg}{\textit{e.g.}\xspace}
\newcommand{\cmark}{\checkmark}

\newcommand{\tool}{AIprint\xspace}    
\newcommand{\tools}{AIprints\xspace} 
\newcommand{\Tool}{AIprint\xspace}    
\newcommand{\Tools}{AIprints\xspace} 

\begin{document}

\title{A Measurement Study of AI-Environment Realism Gaps in Malware-Analysis Sandboxes}


\author{
{\rm Zhiyong Sui, Lamine Noureddine, Mst Eshita Khatun,}\\
{\rm Sideeq Bello, Babangida Bappah, Justin Woodring, Aisha Ali-Gombe}\\
Louisiana State University\\
\{zsui1, lnoureddine, mkhatu3, sbell49, bbappa1, jwoodr7, aaligombe\}@lsu.edu
}

\maketitle


\begin{abstract}
Sandboxing remains a core technique for observing suspicious program behavior, yet environment-aware malware increasingly suppresses execution when analysis is suspected. Prior generations of sandbox evasion focused on virtualization artifacts, timing discrepancies, and wear-and-tear realism. In this paper, we present the first systematic measurement study of AI-environment artifacts as a new sandbox-evasion surface. We operationalize this realism gap through \tools, a probe framework that captures persistent artifacts left behind by AI-capable software ecosystems, including AI-assistant configuration directories, model caches, environment variables, local inference services, and package dependencies.

We systematically extract 450 unique artifacts from 284 open-source AI projects on GitHub, compile them into unprivileged Windows probes, and evaluate them across seven commercial and open-source sandbox backends together with three AI-capable reference hosts. Our results show that traditional VM-detection baselines fail to reliably distinguish real AI-capable systems from modern sandboxes, whereas twelve AI-environment artifacts appear on the reference hosts and on none of the evaluated backends. A controlled 214-step installation experiment establishes a causal relationship between AI tool and package installation and measurable AI-environment artifact accumulation, while adaptive spoofing experiments reveal a fundamental operational asymmetry: reproducing convincing AI software environments is substantially more expensive than detecting shallow spoofing.
\end{abstract}



\section{Introduction}\label{intro}
Malware-analysis sandboxes are a foundational component of modern security infrastructure. Analysts routinely execute untrusted programs inside contained and instrumented environments because static analysis alone is often ineffective against packing, obfuscation, and runtime-triggered behavior~\cite{willems2007cwsandbox, egele2008survey}. The entire containment approach rests on a fragile assumption: that the sandbox resembles a genuine endpoint closely enough to avoid detection. Malware authors have spent more than a decade attacking that assumption. 

Defenders responded in three successive generations: first by scrubbing obvious virtualization artifacts~\cite{petsas2014rage, klein2008scoopyng, vidas2014evading}, then by reducing timing and emulation side channels and deploying bare-metal analysis hosts~\cite{kang2009emulating, bulygin2008cpu}, and most recently by aging sandbox images to mimic the wear and tear of real user activity~\cite{miramirkhani2017spotless}. Each generation focused on reducing observable discrepancies between sandbox environments and genuine endpoints.

Modern malware, however, is increasingly shifting from probing what a host is to probing what the host runs. In early 2025, Google's Threat Intelligence Group disclosed HONESTCUE, malware that invokes the Gemini API to generate and execute payloads in memory~\cite{google2025honestcue}; shortly afterward, Trend Micro documented Lumma Stealer checking for GPU presence before detonation~\cite{lummastealer2025}. At the same time, AI-capable software ecosystems have become increasingly common across consumer, enterprise, and developer environments, creating a new class of persistent host-level footprints that differ fundamentally from traditional virtualization or wear-and-tear indicators.

Thus, in this paper, we present the first systematic measurement study of AI-environment artifacts as a new sandbox-evasion surface. We operationalize this realism gap through \tool, a probe framework that captures persistent artifacts left behind by AI-capable software ecosystems. Unlike earlier anti-analysis heuristics that infer human activity indirectly through behavioral traces such as mouse movement or process counts~\cite{afianian2019malware, alkhaser}, \tool focus on system artifacts associated with functioning AI tooling, including AI-assistant configuration directories, model caches, environment variables, local inference services, and package dependencies. These artifacts emerge as a byproduct of installed AI software and are therefore substantially more expensive to reproduce convincingly than traditional virtualization or wear-and-tear indicators.

To evaluate this AI-environment fidelity gap, we design a four-stage measurement methodology that systematically extracts 450 unique artifacts from 284 open-source AI projects on GitHub, compiles them into standalone unprivileged Windows probes, and executes them across seven commercial and open-source sandbox backends together with three AI-capable hosts. We further establish causality through a controlled cumulative-installation experiment on a clean machine and evaluate defensive spoofing strategies against adaptive structural validation attacks.

Our results show that traditional VM-detection baselines fail to distinguish AI-capable hosts from modern sandboxes, whereas twelve AI-environment artifacts---including \texttt{.claude}, \texttt{.cursor}, \texttt{OPENAI\_API\_KEY}, and Ollama port~11434---appear on the reference hosts and on none of the evaluated backends. Across 214 cumulative installation steps, \tool detection rises monotonically from 0.6\% to 17.8\% while the traditional VM baseline remains effectively unchanged. Finally, we show that shallow artifact spoofing is insufficient against lightweight structural validation, creating a fundamental operational asymmetry in which reproducing convincing AI ecosystems is substantially more expensive than detecting shallow spoofing.

Our paper makes the following main contributions:
\begin{itemize}[leftmargin=1.5em]
    \item We present the first systematic measurement study of AI-environment artifacts as a new sandbox-evasion surface and organize these artifacts into a four-category taxonomy extracted from 284 GitHub AI projects through a reusable provenance-preserving pipeline (\S\ref{sec:artifacts}).
    \item We evaluate \tool across seven commercial and open-source sandbox backends together with three AI-capable hosts, showing that AI-environment artifacts discriminate where traditional VM-detection baselines do not (\S\ref{sec:eval_rq1}, \S\ref{sec:eval_rq2}).
    \item We establish a causal relationship between AI-tool installation and \tool detection through a 214-step controlled cumulative-installation experiment, using traditional VM detection as a negative control (\S\ref{sec:controlled_experiment}).
    \item We evaluate progressive artifact-spoofing defenses and adaptive structural validation attacks across five spoofing levels, exposing a fundamental operational asymmetry in which reproducing convincing AI ecosystems is substantially more expensive than detecting shallow spoofing (\S\ref{sec:eval_rq4}).
    \item To support reproducible research, we will release the complete \tool framework, artifact-extraction workflow, generated datasets, probe implementations, and experimental results upon publication.
\end{itemize}

\noindent \textbf{Paper outline.} The remainder of this paper is organized as follows. Section~\ref{sec:background} presents the background and related work on sandbox evasion and AI-environment realism. Section~\ref{sec:meth} introduces the overall measurement methodology and workflow. Sections~\ref{sec:artifacts} and~\ref{sec:measurement} describe the artifact taxonomy, extraction pipeline, probe design, and sandbox execution infrastructure. Section~\ref{sec:evaluation} presents the empirical evaluation and defense analysis across all four research questions. Section~\ref{sec:discussion} discusses limitations, implications, and broader considerations for sandbox realism, and Section~\ref{sec:conclusion} concludes the paper.


\section{Background and Related Work}
\label{sec:background}

Malware-analysis sandbox evasion has evolved through distinct generations of environment fingerprinting.

\PP{First generation: virtualization artifacts}
Early sandbox-evasion techniques focused on deterministic fingerprints left by virtualization platforms. Malware probed for artifacts such as CPUID vendor strings (e.g., \texttt{VMwareVMware}), MAC-address prefixes, hypervisor-specific drivers and processes, and characteristic registry keys~\cite{petsas2014rage, klein2008scoopyng, vidas2014evading}. Toolkits such as al-khaser aggregate hundreds of these checks into reusable anti-analysis frameworks~\cite{alkhaser}. Defenders responded by scrubbing or spoofing virtualization artifacts within sandbox images.

\PP{Second generation: timing, emulation, and bare metal}
As virtualization artifacts became easier to hide, malware shifted toward timing and behavioral discrepancies between analysis environments and physical hardware, including instruction latency, cache behavior, and imperfect emulation~\cite{kang2009emulating, bulygin2008cpu}. Defenders responded through increasingly realistic emulation, bare-metal execution, and behavioral-divergence analysis to expose ``split-personality'' malware that suppresses malicious behavior during analysis~\cite{kruegel2014fullsystem, moser2007multipath, kirat2014barecloud, lindorfer2011detecting, balzarotti2010split, gorter2023enviral}.

\PP{Third generation: system wear-and-tear}
As virtualization and timing artifacts became less reliable, malware increasingly evaluated whether a system appeared genuinely used rather than merely virtualized. Real endpoints accumulate traces of long-term human activity, including browser history, event logs, user files, and diverse background processes, whereas sandbox images are often freshly reverted and operationally sparse. Anti-analysis heuristics exploit this gap through indicators such as realistic process counts, mouse activity, populated user directories, and anomalous DNS behavior~\cite{afianian2019malware, alkhaser}. Miramirkhani et al.\ formalized this idea as \emph{wear-and-tear fingerprinting}, showing that environmental usage artifacts can distinguish genuine hosts from sandbox environments with 92.86\% accuracy~\cite{miramirkhani2017spotless}. Defenders consequently adopted \emph{synthetic aging} to simulate realistic system evolution and user activity.

\PP{Fingerprinting the sandbox, or the real machine}
A complementary line of work fingerprints sandbox infrastructures directly rather than validating environment characteristics. Systems such as SandPrint, AVLeak, and Android-environment fingerprinting identify analysis environments through high-entropy hardware allocations, emulator side channels, and mobile-environment traces~\cite{yokoyama2016sandprint, blackthorne2016avleak, maier2014divideandconquer, oberheide2012bouncer}.

However, many such indicators remain deterministic and can potentially be diversified or randomized by careful sandbox operators. 

\subsection{A Fourth Generation of Evasion}
We study a gap not addressed by prior generations of sandbox evasion: the presence or absence of persistent AI software ecosystems. Unlike earlier approaches that focused on virtualization artifacts, execution characteristics, or passive traces of human activity, this generation targets evidence of AI-capable computational environments, including local inference runtimes, model caches, package dependencies, runtime services, and AI-specific configuration state. This shift also changes the defender's burden. Prior generations could often be mitigated through artifact scrubbing, higher-fidelity execution, or synthetic aging. In contrast, mitigating AI-environment checks requires sandbox operators to deploy and continuously maintain functional AI ecosystems with evolving runtimes, dependencies, model artifacts, and potentially GPU-enabled infrastructure. Table~\ref{tab:generation_comparison} in Appendix~\ref{ap:gen} summarizes how the four generations differ in the environmental fingerprints they exploit and the operational cost of defending against them.

\subsection{AI-Environment Realism}
\label{sec:ai_realism}
The evasion surface studied in this work is driven by the rapid adoption of AI-enabled software ecosystems across consumer, enterprise, and developer environments. PyTorch averages over 80~million monthly PyPI downloads as of 2025~\cite{pypistats2025torch}; Ollama exceeds 100~million Docker Hub pulls~\cite{ollama2024}; and the JetBrains 2025 Developer Ecosystem Survey reports that 62\% of professional developers use at least one AI coding assistant~\cite{jetbrains2025}. At the enterprise level, GitHub reports that Copilot is deployed across more than 77{,}000 organizations and used by over 1.8~million developers~\cite{github2024copilot}. Beyond software engineering, operating-system vendors increasingly position AI capability as a standard endpoint feature through platforms such as Copilot+ PCs, Apple Intelligence, and Gemini-integrated productivity environments. These ecosystems leave behind persistent local artifacts, including model caches, runtime services, package dependencies, environment variables, AI assistants, and configuration directories that remain observable long after active use has ended. We refer to the resulting environmental characteristics as \emph{AI-environment realism}.

We conceptualize AI-enabled endpoints as a spectrum of \emph{AI personas}, ranging from lightweight consumer AI integrations and enterprise AI-assisted workflows to fully provisioned local AI ecosystems. These personas collectively contribute to a widening fidelity gap between genuine endpoints and malware-analysis sandboxes. Three properties make AI-environment realism particularly relevant to sandbox evasion. First, \textbf{ubiquity}: AI tooling increasingly spans consumer, enterprise, and developer workflows, causing modern systems to accumulate artifacts from multiple AI subsystems. Second, \textbf{persistence}: artifacts such as model caches, runtime services, package dependencies, and AI-specific configuration directories survive reboots and remain observable long after active use has ended. Third, \textbf{structural absence from sandboxes}: malware-analysis environments are optimized for scalability and execution fidelity rather than reproducing genuine AI workflows, making convincing AI ecosystems operationally expensive to deploy and maintain. While all AI personas contribute to this ecosystem gap, local AI personas expose the richest and most externally observable artifact surface through installed runtimes, inference frameworks, GPU tooling, package managers, and evolving configuration state. Accordingly, this paper explores two motivating questions: \textit{Can AI-capable software ecosystems create observable differences between genuine endpoints and malware-analysis sandboxes? If so, do these differences introduce a practical new surface for sandbox evasion and defensive adaptation?}

\subsection{Threat Model}
\label{sec:threat_model}
We model a malware sample executing as an unprivileged user-mode process on a Windows host. The adversary can invoke standard Win32 APIs accessible to normal user processes (e.g., \texttt{PathFileExists}, \texttt{GetEnvironmentVariable}, and \texttt{connect})~\cite{msdn2024pathfileexists, msdn2024getenv}, but has no kernel-mode access, administrator privileges, or ability to install additional software.

We consider two adversarial objectives. \textbf{Adversary~A (AI-infrastructure reconnaissance)} probes compromised hosts for locally accessible AI resources, including inference servers, exposed API credentials, GPU-enabled runtimes, or installed AI tooling that may support payload generation or post-compromise automation~\cite{google2025honestcue, google2025promptflux}. Prior work has shown that environmental state itself can influence malware execution behavior~\cite{10.1145/2046707.2046740}. Recent malware already demonstrates related behavior: earlier work documented GPU-presence checks in malicious documents~\cite{desai2016antivmsandbox}, HONESTCUE invokes external Gemini APIs during execution, and Lumma Stealer checks for GPU presence prior to detonation~\cite{google2025honestcue, lummastealer2025}. \textbf{Adversary~B (sandbox evasion)} interprets the absence of persistent AI-environment artifacts as evidence of sandbox execution. Environments lacking indicators associated with AI-capable workflows---such as model caches, inference services, package dependencies, or AI-assistant directories (e.g., \texttt{.claude/}, \texttt{.cursor/})---are treated as more likely to be analysis systems than genuine endpoints. 

Our evaluation focuses exclusively on Adversary~B. Adversary~A is included only to motivate the broader relevance of AI-environment artifacts in modern computing environments.

\begin{figure*}[h]
\centering
\begin{tikzpicture}[
    box/.style={rectangle, draw, rounded corners=2pt, minimum width=1.4cm,
                minimum height=0.55cm, align=center, font=\scriptsize},
    grp/.style={draw, dashed, rounded corners=4pt, fill=#1,
                inner xsep=7pt, inner ysep=5pt,
                minimum height=2.8cm},          
    arr/.style={->, >=Stealth, semithick},
    stitle/.style={font=\scriptsize\bfseries},
    note/.style={font=\tiny, text=gray!60!black, align=center},
    sref/.style={font=\tiny\itshape, text=black!55}
]

\coordinate (TOP) at (0, 1.5);
\coordinate (BOT) at (0,-1.7);

\node[box, fill=blue!10]  at ( 0, 1.2)  (gh)     {GitHub AI\\Projects};
\node[note, below=1pt of gh]                      {284 repos};
\node[box, fill=blue!10]  at ( 0, 0.0)  (parse)  {Parse\\Configs};
\node[box, fill=blue!10]  at ( 0,-1.2) (artset) {Artifact\\Set};
\node[note, below=1pt of artset]                   {450 unique};
\draw[arr] (gh) -- (parse);
\draw[arr] (parse) -- (artset);
\begin{scope}[on background layer]
  \node[grp=blue!4, fit=(gh)(artset)(BOT -| gh)] (S1) {};
\end{scope}
\node[stitle, above=2pt of S1.north] {Stage 1: Extract};
\node[sref,   below=3pt of S1.south] (s1r) {\S\ref{sec:github_extraction}};

\node[box, fill=orange!12] at (3.2, 0.0) (comp) {Compile\\Probes};
\node[note, below=1pt of comp] {Win EXE};
\draw[arr] (artset.east) -- (comp.west);
\begin{scope}[on background layer]
  \node[grp=orange!4, fit=(TOP -| comp)(comp)(BOT -| comp)] (S2) {};
\end{scope}
\node[stitle, above=2pt of S2.north] {Stage 2: Probe};
\node[sref,   below=3pt of S2.south] (s2r) {\S\ref{sec:implementation}};

\node[box, fill=red!10]    at (6.0, 0.45)  (sb)   {Sandbox\\Backends};
\node[box, fill=teal!12]   at (6.0,-0.45)  (rh)   {Reference\\Host};
\node[box, fill=purple!10] at (8.0, 0.0)   (coll) {Collect \&\\Compare};
\node[note, below=1pt of coll] {DNS exfil};
\node[box, fill=gray!12]   at (8.0,-1.3)  (cls)  {Classify};

\draw[arr] (comp.east) -- ++(0.5,0) |- (sb.west);
\draw[arr] (comp.east) -- ++(0.5,0) |- (rh.west);
\draw[arr] (sb)  -- (coll);
\draw[arr] (rh)  -- (coll);
\draw[arr] (coll) -- (cls);

\begin{scope}[on background layer]
  \node[grp=red!4, fit=(TOP -| sb)(sb)(rh)(coll)(cls)(BOT -| cls)] (S3) {};
\end{scope}
\node[stitle, above=2pt of S3.north] {Stage 3: Evaluate};
\node[sref,   below=3pt of S3.south] (s3r) {\S\ref{sec:evaluation}};

\node[box, fill=green!10] at (10.5, 1.2)  (cln)  {Clean\\Machine};
\node[box, fill=green!10] at (10.5, 0.0)  (inst) {Cumulative\\AI Install};
\node[box, fill=green!18] at (10.5,-1.2) (crv)  {Growth\\Curve};
\draw[arr] (cln) -- (inst);
\draw[arr] (inst) -- (crv);
\begin{scope}[on background layer]
  \node[grp=green!4, fit=(cln)(crv)(BOT -| crv)] (S4) {};
\end{scope}
\node[stitle, above=2pt of S4.north] {Stage 4: Validate};
\node[sref,   below=3pt of S4.south] (s4r) {\S\ref{sec:controlled_experiment}};

\draw[arr] (cls.east) -- ++(0.3,0) |- (cln.west);

\end{tikzpicture}
\caption{End-to-end methodology.
\textbf{Stage~1}: extract artifacts from 284 GitHub AI projects (\S\ref{sec:github_extraction}).
\textbf{Stage~2}: compile into unprivileged Win32 probes (\S\ref{sec:implementation}).
\textbf{Stage~3}: deploy to seven sandboxes and one reference host; classify artifacts into three discrimination classes (\S\ref{sec:evaluation}).
\textbf{Stage~4}: validate causality via controlled installation on a separate clean machine (\S\ref{sec:controlled_experiment}).
RQ4 (defense evaluation) is addressed in \S\ref{sec:eval_rq4}.}
\label{fig:workflow}
\end{figure*}
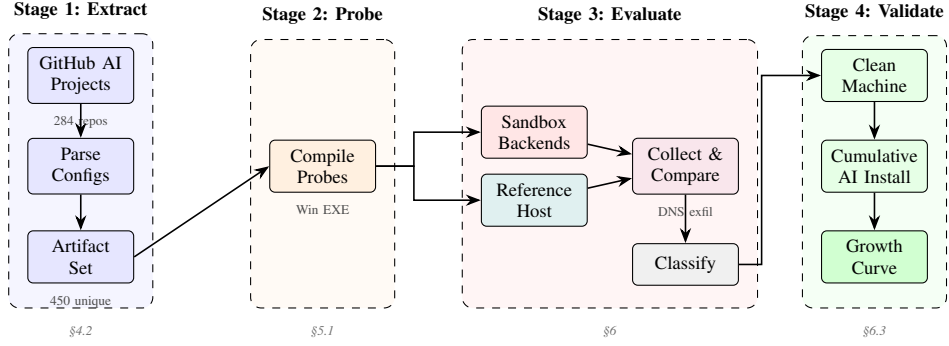

\section{Methodology Overview}
\label{sec:meth}

Our methodology formalizes AI-environment realism as a measurable artifact surface and evaluates whether those artifacts systematically distinguish genuine AI-capable endpoints from malware-analysis sandboxes. Formally, let $A=\{a_1,a_2,\dots,a_n\}$ denote the extracted AI-environment artifact set, where each artifact corresponds to a lightweight user-mode check (e.g., directory presence, environment variable, listening port, or installed package). Given an execution environment $E$, each probe evaluates an artifact-presence function $f(a_i,E)\in\{0,1\}$ indicating whether artifact $a_i$ is observable within $E$. Our objective is to measure how artifact distributions differ across AI-capable endpoints and malware-analysis sandboxes, and to identify artifact subsets that exhibit strong cross-environment discrimination behavior.

Our proposed workflow operates in four stages (Figure~\ref{fig:workflow}). \textbf{Stage~1 (Extract)} systematically harvests checkable artifacts from 284 GitHub AI projects by parsing configuration files and extracting persistent AI-environment indicators (\S\ref{sec:github_extraction}).
\textbf{Stage~2 (Probe)} compiles these artifacts into standalone, unprivileged Windows executables that perform artifact checks and report results via DNS-based exfiltration (\S\ref{sec:implementation}--\S\ref{sec:dns_exfil}).
\textbf{Stage~3 (Evaluate)} deploys probes across seven commercial and open-source sandbox backends and three reference hosts, then measures artifact discrimination behavior across sandbox and reference environments (RQ1--RQ2, \S\ref{sec:eval_rq1}--\S\ref{sec:eval_rq2}).
\textbf{Stage~4 (Validate)} evaluates causality through controlled cumulative AI-software installation on a separate clean machine and assesses defensive spoofing strategies (RQ3--RQ4, \S\ref{sec:controlled_experiment}, \S\ref{sec:eval_rq4}).

\section{Artifact Corpus Construction}
\label{sec:artifacts}

\begin{table*}[t]
\centering
\caption{Artifact taxonomy. Counts shown are for the Trending-41 dataset; the Topics-253 dataset yields substantially larger artifact sets under the same extraction pipeline.} 
\label{tab:artifact_taxonomy}
\small
\setlength{\tabcolsep}{4pt}
\begin{tabular}{@{}
  >{\raggedright\arraybackslash}p{1.5cm}
  >{\raggedright\arraybackslash}p{4.8cm}
  >{\raggedright\arraybackslash}p{3.0cm}
  >{\raggedright\arraybackslash}p{4.0cm}@{}}
\toprule
\textbf{Category} & \textbf{Representative Artifacts} & \textbf{Main Detection API} & \textbf{Representative Source Tools} \\
\midrule
\textbf{Directories}\newline(19) &
  \texttt{.claude}, \texttt{.cursor}, \texttt{.copilot}, \texttt{.ollama}, \texttt{.cache/huggingface}, \texttt{.conda}, \texttt{.jupyter} &
  \texttt{PathFileExists()} &
  Claude~\cite{anthropic2024claude}, Cursor, Copilot, Ollama~\cite{ollama2024}, HuggingFace~\cite{huggingface2024hub} \\
\midrule
\textbf{Env.\ Vars}\newline(144) &
  \texttt{OPENAI\_API\_KEY}, \texttt{ANTHROPIC\_API\_KEY}, \texttt{HF\_TOKEN}, \texttt{OLLAMA\_HOST}, \texttt{CUDA\_PATH} &
  \texttt{GetEnvironment-}\newline\texttt{Variable()} &
  OpenAI~\cite{openai2024api}, Anthropic~\cite{anthropic2024claude}, CUDA~\cite{nvidia2024cuda}, Conda~\cite{anaconda2024} \\
\midrule
\textbf{Ports}\newline(28) &
  11434 (Ollama), 1234 (LM~Studio), 7860 (Gradio), 8888 (Jupyter) &
  \texttt{socket()} +\newline\texttt{connect()} &
  Ollama~\cite{ollama2024}, LM~Studio~\cite{lmstudio2024}, Gradio~\cite{gradio2024}, Jupyter~\cite{jupyter2024} \\
\midrule
\textbf{Packages}\newline(264) &
  \texttt{torch}, \texttt{tensorflow}, \texttt{transformers}, \texttt{langchain}, \texttt{openai}, \texttt{chromadb} &
  \texttt{PathFileExists()}\newline on \texttt{site-packages/} &
  PyTorch~\cite{paszke2019pytorch}, HuggingFace~\cite{wolf2020transformers}, LangChain~\cite{langchain2024} \\
\bottomrule
\end{tabular}
\end{table*}

\subsection{Artifact Taxonomy}
\label{sec:artifact_categories}
Prior sandbox-evasion research organizes checks by detection mechanism: VM artifacts probe hardware identity, timing checks measure execution speed, and wear-and-tear checks inspect accumulated usage~\cite{miramirkhani2017spotless, yokoyama2016sandprint}.
We adopt a complementary organization by artifact type, selecting categories that satisfy three criteria derived from the threat model (\S\ref{sec:threat_model}): (C1)~low execution cost, (C2)~no elevated privileges, and (C3)~API behavior consistent with legitimate application startup behavior
These criteria exclude, for example, WMI queries (which incur multi-second latency and may be monitored) and kernel-mode checks (which violate~C2).

We identify four artifact categories designed to satisfy all three criteria (Table~\ref{tab:artifact_taxonomy}). Their runtime and implementation characteristics are evaluated during the probing phase (\S\ref{sec:implementation}. All four categories share a property that distinguishes them from prior evasion generations: the underlying APIs (\texttt{PathFileExists}, \texttt{GetEnvironmentVariable}, and \texttt{connect}) are routinely invoked during legitimate application startup and require no elevated privileges. In contrast, prior generations often relied on specialized instructions or monitoring-sensitive operations, such as CPUID checks, RDTSC timing loops, or WMI queries~\cite{yokoyama2016sandprint}. 

\subsection{GitHub Extraction Pipeline}
\label{sec:github_extraction}
Rather than enumerating AI-environment artifacts manually, we construct the artifact corpus through an automated extraction pipeline that analyzes configuration files from real-world AI ecosystem projects hosted on GitHub. The pipeline preserves provenance by mapping each extracted artifact back to a specific repository and source file.

\PP{Project Collection}
We collect projects from two complementary GitHub sources. First, we crawl repositories associated with high-popularity AI-related topics (\texttt{ai}, \texttt{llm}, \texttt{mcp}), yielding 253 Windows-compatible projects after deduplication and filtering. Second, we collect 41 actively maintained projects from GitHub Trending using AI-focused keyword filtering. The combined dataset contains 284 unique projects and 4{,}576 extracted artifact checks. Throughout the paper, the Trending-41 dataset provides interpretable per-project analysis, while the larger Topics-253 dataset demonstrates extraction scalability.

\PP{Artifact Extraction}
For each project, we parse configuration-related files through GitHub's raw content API, including README files, Python dependency manifests, and Docker configuration files. README files provide broad but weakly structured artifact coverage, while dependency manifests and Docker files expose authoritative package, environment-variable, directory, and port information. The extraction pipeline identifies artifact references, normalizes them into the taxonomy defined in \S\ref{sec:artifact_categories}, and preserves repository-level provenance for all extracted entries.

\subsection{Artifact Filtering and Normalization}
Raw extraction yields 197 environment variables and 325 Python packages. We normalize the corpus through two filtering steps. First, we remove environment variables present in a stock Windows~11 installation (e.g., \texttt{PATH}, \texttt{HOME}, \texttt{COMPUTERNAME}) to eliminate generic operating-system artifacts. Second, we exclude Python packages associated with general-purpose software development rather than AI/ML workflows (e.g., \texttt{pytest}, \texttt{setuptools}, \texttt{wheel}) using PyPI package metadata and classifier information. After filtering and cross-category deduplication, the final artifact corpus contains 450 unique artifact mappings across four categories: 19 directories, 144 environment variables, 28 network ports, and 264 Python packages. We intentionally preserve broad artifact coverage at this stage; subsequent evaluation identifies the subset that exhibits meaningful cross-environment discrimination behavior.

\begin{table}[t]
\centering
\caption{Top contributing Trending AI projects by artifact count (6 of 41 shown). Total 455 raw mappings across 41 projects; 450 unique after deduplication.}
\label{tab:project_artifact_summary}
\small
\begin{tabular}{@{}lccccc@{}}
\toprule
\textbf{Project} & \textbf{Dirs} & \textbf{Env} & \textbf{Ports} & \textbf{Pkgs} & \textbf{Total} \\
\midrule
vxcontrol/pentagi & 3 & 62 & 9 & 46 & 120 \\
rowboatlabs/rowboat & 1 & 16 & 5 & 29 & 51 \\
BerriAI/litellm & 0 & 12 & 3 & 22 & 37 \\
ollama/ollama & 1 & 3 & 1 & 2 & 7 \\
nvidia/cuda-toolkit & 1 & 2 & 0 & 2 & 5 \\
anthropics/claude-code & 1 & 2 & 0 & 1 & 4 \\
\midrule
\textbf{Total (41 projects)} & \textbf{19} & \textbf{144} & \textbf{28} & \textbf{264} & \textbf{455} \\
\bottomrule
\end{tabular}
\end{table}

\subsection{Artifact Clustering}
\label{sec:artifact_clustering}
While individual artifact checks establish presence or absence, grouping artifacts by source project enables evaluation of whether a host exhibits a coherent AI ecosystem rather than isolated traces as shown in Appendix Algorithm~\ref{ap:appendix_algorithm}. For each project $P$, we compute a confidence score:

\[
\text{confidence}(P)=\frac{|\text{detected}(P)|}{|\text{total}(P)|}
\]

A project is considered ``detected'' if its confidence exceeds 15\%, allowing small projects to trigger on a minimal number of matching artifacts while requiring broader agreement for larger artifact sets. Because popular artifacts (e.g., \texttt{.claude}, \texttt{OPENAI\_API\_KEY}) appear across multiple projects, project-level scores should be interpreted as indicators of AI-ecosystem realism rather than precise evidence of individual project installation. Consequently, clustering serves as supporting evidence of environment-level AI capability, while our primary analysis focuses on artifact-level cross-environment discrimination behavior.
\section{Probe Design and Measurement Infrastructure}
\label{sec:measurement}
Having constructed the AI-environment artifact corpus (\S\ref{sec:artifacts}), we now describe \tool, the measurement infrastructure that implements each artifact as a lightweight user-mode check and collects results from constrained sandbox environments. Our design emphasizes low-privilege execution, minimal behavioral footprint, and compatibility with sandbox environments that restrict conventional outbound communication, while remaining consistent with the low-cost and user-mode design constraints defined in criteria C1--C3 (\S\ref{sec:artifact_categories}).

\subsection{Probe Architecture}
\label{sec:implementation}
We implement \tool probes as standalone C++ Windows executables compiled with MSVC (Visual Studio 2022) in both 32-bit and 64-bit PE formats to maximize compatibility across heterogeneous sandbox environments. Each artifact corpus (Trending-41 and Topics-253) supports two execution modes: (1)~project-level probes that compute aggregate artifact confidence scores and (2)~single-artifact probes that report individual check outcomes.

Each artifact category maps to a lightweight Win32 API operation. \textbf{1. Directory checks} invoke \texttt{PathFileExistsW()} on expected AI-tool paths expanded from user-specific directories (e.g., \texttt{\%USERPROFILE\%}, \texttt{\%APPDATA\%}). \textbf{2. Environment-variable checks} invoke \texttt{GetEnvironmentVariableW()} and test only for variable presence without inspecting contents. \textbf{3. Port checks} attempt local TCP connections against expected inference-service ports. \textbf{4. Python-package checks} verify package presence within detected \texttt{site-packages/} directories. All checks execute entirely in user mode without requiring administrator privileges or UAC prompts, maintain lightweight execution behavior, and closely resemble legitimate application startup activity, thereby satisfying criteria C1--C3. Because commercial sandboxes frequently cache binaries by cryptographic hash, we generate multiple functionally equivalent probe variants through non-semantic PE modifications to ensure independent analysis across repeated submissions. Probe outputs are serialized into a compact structured format containing artifact identifiers, categories, and check outcomes before being forwarded through the DNS exfiltration pipeline described in \S\ref{sec:dns_exfil}.

\subsection{DNS Exfiltration}
\label{sec:dns_exfil}
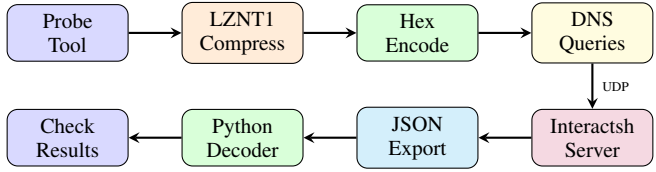
\begin{figure}[t]
\centering
\begin{tikzpicture}[
    node distance=0.7cm,
    box/.style={rectangle, draw, rounded corners, minimum width=1.6cm, minimum height=0.6cm, align=center, font=\footnotesize},
    arrow/.style={->, >=stealth, thick},
    label/.style={font=\tiny, align=center}
]
\node[box, fill=blue!15] (probe) {Probe\\Tool};
\node[box, fill=orange!15, right=of probe] (compress) {LZNT1\\Compress};
\node[box, fill=green!15, right=of compress] (encode) {Hex\\Encode};
\node[box, fill=yellow!15, right=of encode] (dns) {DNS\\Queries};
\node[box, fill=purple!15, below=0.6 cm of dns] (interactsh) {Interactsh\\Server};
\node[box, fill=cyan!15, left=of interactsh] (json) {JSON\\Export};
\node[box, fill=green!15, left=of json] (decode) {Python\\Decoder};
\node[box, fill=blue!15, left=of decode] (results) {Check\\Results};
\draw[arrow] (probe) -- (compress);
\draw[arrow] (compress) -- (encode);
\draw[arrow] (encode) -- (dns);
\draw[arrow] (dns) -- node[right, label] {UDP} (interactsh);
\draw[arrow] (interactsh) -- (json);
\draw[arrow] (json) -- (decode);
\draw[arrow] (decode) -- (results);
\end{tikzpicture}
\caption{DNS exfiltration pipeline. The probe compresses results, hex-encodes them into DNS subdomain labels, and transmits via UDP; an offline decoder reassembles the payload.}
\label{fig:dns_flow}
\end{figure}

Commercial sandbox environments commonly restrict conventional outbound communication channels such as HTTP~\cite{aws2024sandboxnetwork}. To ensure reliable result collection across heterogeneous backends, our measurement pipeline uses DNS-based exfiltration, leveraging the fact that sandbox environments typically permit DNS resolution for normal network operation. DNS exfiltration is a well-established technique documented in MITRE ATT\&CK (T1048)~\cite{mitre2024dnsexfil, nadler2019dnsexfil}. Figure~\ref{fig:dns_flow} illustrates the pipeline. \tool probe outputs are compressed using Windows' native LZNT1 compression routines, encoded into DNS-compatible subdomain labels, and transmitted to an Interactsh~\cite{interactsh} collection server through a sequence of DNS queries. An offline decoder reconstructs and decompresses the transmitted payload to recover structured probe results. Each execution session is associated with a unique identifier to support reliable reassembly across multi-query transmissions. Full protocol details are provided in Appendix~\ref{ap:dns}.

\subsection{Sandbox Execution Pipeline}
\label{sec:sandbox_collection}
We deploy probes across seven sandbox platforms spanning commercial, open-source, interactive, and multi-backend malware-analysis infrastructures (Table~\ref{tab:sandbox_platforms}). These platforms collectively represent the dominant deployment models used in automated malware analysis, including analyst-oriented triage services, enterprise threat-intelligence systems, and open-source COTS deployments.

\PP{Probe Variants}
We execute \textbf{five probe variants} across all platforms and three reference AI-capable hosts: (Variant~1)~a traditional sandbox-evasion baseline (SOTA) composed of 269 checks from prior anti-analysis frameworks, including al-khaser~\cite{alkhaser}, pafish, VMDE, sems, and CheckPlease. (Variant~2)~Topics-253 project-level probes, (Variant~3)~Topics-253 single-artifact probes (4{,}150 individual artifact checks). (Variant~4)~Trending-41 project-level probes, and (Variant~5)~Trending-41 single-artifact probes (450 individual artifact checks).



\begin{table}[t]
\centering
\caption{Sandbox platforms used in the measurement pipeline. All platforms accept PE executables through public submission interfaces and execute \tool probes within Windows guest environments.}
\label{tab:sandbox_platforms}
\small
\setlength{\tabcolsep}{3pt}
\begin{tabular}{@{}llll@{}}
\toprule
\textbf{Platform} & \textbf{Type} & \textbf{Guest OS} & \textbf{Backends} \\
\midrule
ANY.RUN~\cite{anyrun} & Commercial & Win 10 (x64) & 1 \\
CAPEv2~\cite{capev2} & Open-source & Win 10 (x64) & 1 \\
Cuckoo~\cite{Cuckoo_sandbox} & Open-source & Unknown & 2 \\
Hybrid Analysis~\cite{hybrid-analysis} & Commercial & Win 10 (x64) & 1 \\
Kaspersky~\cite{Kaspersky_opentip} & Commercial & Unknown & 1 \\
Triage~\cite{Triage} & Commercial & Win 10 (x64) & 2 \\
VirusTotal~\cite{virustotal_upload} & Multi-engine & Various & 7 \\
\bottomrule
\end{tabular}
\end{table}

VirusTotal is architecturally distinct because submissions may be routed to multiple independent backend sandboxes executing different guest images. Accordingly, we treat VirusTotal backends as separate analysis environments throughout the evaluation.

\PP{Reference AI-Capable Hosts}
For comparison, we execute identical probes on three Windows~11 AI-capable reference hosts representing distinct local AI environment profiles. The hosts share the same hardware configuration (a 24-core (13th Gen Intel Core i9-13900KF) CPU, 64 GB memory, and an NVIDIA GeForce RTX 4090 GPU (24 GB VRAM)) but differ in their installed AI-tool ecosystems and resulting artifact density. They serve as proof-of-concept reference environments for AI-capable endpoints rather than as a population baseline (\S\ref{sec:discussion}). Each probe variant is executed once per reference host and submitted once per sandbox backend, yielding a point-in-time measurement snapshot (February~2026).

\section{Evaluation}
\label{sec:evaluation}
Our evaluation investigates whether AI-environment artifacts introduce observable differences between contemporary malware-analysis sandboxes and AI-capable endpoints, and whether that gap can be exploited or mitigated in practice. We structure the evaluation around four research questions:

\begin{itemize}[leftmargin=1.2em,itemsep=2pt]
    \item \textbf{RQ1:} Do artifacts associated with the \emph{local AI persona} expose a measurable realism gap between real-world AI-capable systems and contemporary sandbox infrastructures?    
    \item \textbf{RQ2:} Which classes of AI-environment artifacts contribute most strongly to cross-environment discrimination?   
    \item \textbf{RQ3:} Do observed artifact differences arise causally from the installation and accumulation of AI software ecosystems?    
    \item \textbf{RQ4:} Can these AI-environment artifacts be reliably mitigated through defensive spoofing or synthetic AI-environment provisioning?
\end{itemize}

For each research question, we analyze results produced by our end-to-end measurement pipeline and discuss their implications for AI-environment-aware evasion.

\subsection{RQ1: Do \Tools Discriminate Where Traditional VM Checks Fail?}
\label{sec:eval_rq1}
To determine whether AI-environment artifacts expose measurable distinction between genuine endpoints and sandbox environments, we first establish a baseline using \tool probe Variant~1. Table~\ref{tab:sota_baseline} shows that traditional VM checks produce uniformly low trigger rates across all environments (1.9--7.6\%). The reference hosts exhibit trigger rates ranging from 7.4\%--9.9\% (avg.~9.1\%), providing only weak and inconsistent host-versus-sandbox discrimination. Many triggered indicators correspond to Hyper-V features enabled by default in modern Windows~11 systems, illustrating how traditional VM-detection fingerprints have become increasingly noisy and weakly discriminative in contemporary environments. In contrast, \tools produce substantially clearer distinction at both the individual-artifact and project levels. Using the Trending-41 corpus, the reference hosts exhibit higher individual-artifact detection rates than every evaluated sandbox backend, while project-level clustering identifies substantially stronger AI-environment signals on the hosts than on six of the seven sandbox platforms (Table~\ref{tab:ai_era_combined}). These results indicate that AI-environment artifacts expose measurable environmental differences that remain observable across diverse contemporary sandbox infrastructures.

\begin{table}[h]
\centering
\caption{\tool probe Variant~1 -- SOTA baseline: 269~checks from 6 evasion projects.}
\footnotemark
\label{tab:sota_baseline}
\small
\begin{tabular}{@{}lcc@{}}
\toprule
\textbf{Environment} & \textbf{Triggered / 269} & \textbf{Rate} \\
\midrule
ANY.RUN & 5 & 1.9\% \\
Kaspersky & 8 & 3.0\% \\
Cuckoo (v1) & 8 & 3.0\% \\
Triage (v0) & 11 & 4.1\% \\
Triage (v1) & 11 & 4.1\% \\
Hybrid Analysis & 12 & 4.4\% \\
CAPEv2 & 15 & 5.5\% \\
VirusTotal & 8--19 & 3.0\%--7.0\% \\
Cuckoo (v0) & 21 & 7.6\% \\
\midrule
\textbf{Reference Hosts} & \textbf{avg. 24.6} & \textbf{avg. 9.1\%s} \\
\bottomrule
\end{tabular}
\end{table}
\footnotetext{VMDE dynamically adds 1--7 entries when it detects VM artifacts; actual counts range 269--276. Rates are computed against the 269~base checks for consistency across all environments.}

\subsubsection{Baseline Choice}
We compare against traditional VM detection rather than wear-and-tear fingerprinting~\cite{miramirkhani2017spotless} because reproducing the latter would require a large-scale multi-user endpoint study beyond the scope of this work. AI-environment artifacts are conceptually orthogonal to wear-and-tear realism and could potentially complement existing realism-based classifiers.

\summarybox{Traditional VM-detection checks produce uniformly low and weakly discriminative trigger rates across both sandbox and real environments. In contrast, AI-environment artifacts expose substantially stronger separation between contemporary malware-analysis sandboxes and AI-capable endpoints at both the individual-artifact and project levels.}

\subsection{RQ2: Which Artifacts Discriminate?}
\label{sec:eval_rq2}
Having established that AI-environment artifacts expose a measurable distinction, we now identify which artifacts contribute to cross-environment discrimination and whether those artifacts appear consistently across sandbox backends. As shown in Table~\ref{tab:ai_era_combined}, the reference AI-capable hosts consistently exhibit substantially higher project-level and individual-artifact detection rates than the evaluated sandbox backends, indicating that contemporary malware-analysis environments contain markedly sparser AI-environment footprints than real-world AI-capable systems. Detection rates also vary across the reference hosts themselves, reflecting differences in installed AI-tool ecosystems and resulting artifact density.

\begin{table*}[h]
\centering
\caption{\tool detection across all evaluated environments. Six of seven sandbox platforms detect $\leq$7.3\% of Trending-41 projects, while the reference AI-capable hosts consistently exhibit the highest project-level and individual-artifact detection rates. ``---'' denotes data not collected.}
\label{tab:ai_era_combined}
\small
\begin{tabular}{@{}lcccc@{}}
\toprule
 & \multicolumn{2}{c}{\textbf{Topics 253}} & \multicolumn{2}{c}{\textbf{Trending 41}} \\
\cmidrule(lr){2-3} \cmidrule(lr){4-5}
\textbf{Environment} & \textbf{Projects} & \textbf{Single} & \textbf{Projects} & \textbf{Single} \\
\midrule
ANY.RUN & 0.4\% & 0.05\% & 0.0\% & 0.0\% \\
Kaspersky & 0.4\% & 0.05\% & 0.0\% & 0.0\% \\
Triage & 0.4\% & 0.05\% & 0.0\% & 0.0\% \\
Cuckoo & 0.1\% & 0.07\% & 2.4\% & 0.2\% \\
CAPEv2 & 11.5\% & 0.07\% & 7.3\% & 0.2\% \\
Hybrid Analysis & 29.6\% & 0.34\% & 0.0\% & 1.3\% \\
VirusTotal (range) & 0.4--11.9\% & 0.05--0.36\% & 0.0--26.8\% & 0.0--1.3\% \\
\midrule
Real Host 1 & 33.2\% & 0.41\% & 46.3\% & 1.8\% \\
Real Host 2 & \textbf{53.0\%} & \textbf{2.4\%} & \textbf{56.1\%} & \textbf{10.0\%} \\
Real Host 3 & 45.1\% & 0.9\% & 34.1\% & 3.3\% \\
\midrule
\textbf{Discrimination Gap} & +3.6--52.9\% & +0.05--2.35\% & \textbf{+7.3--56.1\%} & +0.5--10.0\% \\
\bottomrule
\end{tabular}
\end{table*}

\subsubsection{Cross-Sandbox Detection Rates}
We observe that detection rates vary sharply across the sandbox backends. Five of seven sandbox platforms detect $\leq$7.3\% of Trending-41 projects, while one VirusTotal backend instance reaches 26.8\%, illustrating substantial cross-backend heterogeneity. Hybrid Analysis exhibits an elevated project-level rate (29.6\%) due to the presence of generic Python ecosystem artifacts that inflate clustering scores despite the absence of AI-specific tooling. Consistent with this interpretation, Hybrid Analysis triggers only 0.34\% of individual artifact checks.

\subsubsection{Three Artifact Classes}
\label{sec:three_classes}
Of the 4{,}150 individual artifacts probed, only 27 triggered in all evaluated environments. We partition these artifacts into three disjoint classes based on their cross-environment behavior (Table~\ref{tab:three_classes}). \textbf{Class~C} forms the primary discriminating surface: twelve artifacts appear exclusively on the reference AI-capable hosts and are absent from every evaluated sandbox backend. These artifacts are uniformly AI-specific and include AI-assistant directories (\texttt{.cursor}, \texttt{.claude}, \texttt{.copilot}, \texttt{.gemini}), the HuggingFace model cache, the \texttt{OPENAI\_API\_KEY} environment variable, Ollama's local inference port~11434, and additional AI-runtime configuration directories.

\begin{table}[t]
\centering
\caption{Derived artifact classes.} 
\label{tab:three_classes}
\small
\begin{tabular}{@{}lcl@{}}
\toprule
\textbf{Class} & \textbf{Count} & \textbf{Composition} \\
\midrule
A: Sandbox-resident & 10 & Generic Python pkgs (Hybrid only) \\
B: Shared & 5 & OS env vars (\texttt{PATH}, \texttt{HOMEPATH}) \\
C: Host-exclusive & 12 & AI IDE dirs, ML caches, \\
   &    & API keys, LLM ports \\
\bottomrule
\end{tabular}
\end{table}

\subsubsection{Discriminating Artifacts}
Table~\ref{tab:unique_artifacts} lists all twelve Class~C artifacts observed exclusively on the reference AI-capable hosts and absent from every evaluated sandbox backend. We noted that the Class~C set spans multiple independent layers of the local AI ecosystem, including AI assistants, local inference runtimes, model caches, GPU tooling, notebook infrastructure, and API credentials, suggesting that the observed distinction is ecosystem-level rather than tied to any single tool category.

\begin{table}[t]
\centering
\caption{Class~C artifacts exhibiting strong host-versus-sandbox separation.}
\label{tab:unique_artifacts}
\small
\begin{tabular}{@{}lll@{}}
\toprule
\textbf{Category} & \textbf{Artifact} & \textbf{Source Tool} \\
\midrule
AI IDE Config & \texttt{\textasciitilde{}/.cursor} & Cursor AI \\
AI IDE Config & \texttt{\textasciitilde{}/.claude} & Claude Code \\
AI IDE Config & \texttt{\textasciitilde{}/.copilot} & GitHub Copilot \\
AI IDE Config & \texttt{\textasciitilde{}/.gemini} & Gemini CLI \\
ML Cache & \texttt{.cache/huggingface} & HuggingFace Hub \\
LLM Data & \texttt{\textasciitilde{}/.ollama} & Ollama \\
Python Env & \texttt{\textasciitilde{}/.conda} & Anaconda \\
Notebooks & \texttt{\textasciitilde{}/.jupyter} & Jupyter \\
GPU Runtime & \texttt{CUDA\_PATH} dir & NVIDIA CUDA \\
API Key & \texttt{OPENAI\_API\_KEY} & OpenAI SDK \\
Local LLM & Port 11434 & Ollama server \\
Notebooks & Port 8888 & Jupyter server \\
\bottomrule
\end{tabular}
\end{table}

\subsubsection{Per-Category Analysis}
Artifact discrimination varies substantially across categories. \textbf{Directories} provide the strongest separation: nine AI-specific directories appear on the reference AI-capable hosts and none on any evaluated sandbox backend. \textbf{Ports} also exhibit strong distinction, with Ollama (11434) and Jupyter (8888) services observed only on the reference hosts. \textbf{Environment variables} produce mixed signal: AI-specific credentials such as \texttt{OPENAI\_API\_KEY} are host-exclusive, whereas generic variables (e.g., \texttt{PATH}) appear across all environments. \textbf{Python packages} discriminate least strongly because some sandbox images ship generic Python ecosystems that partially overlap with AI-oriented dependency stacks. Additional project-level clustering results, threshold-sensitivity analysis, and VirusTotal backend heterogeneity measurements are provided in Appendix~\ref{sec:appendix_rq2_details}. These additional experiments reproduce the same asymmetry observed in the main results: AI-capable reference environments consistently exhibit substantially richer AI-environment artifact presence than contemporary sandbox backends.

\summarybox{
Twelve Class~C artifacts appear on the reference AI-capable hosts only, creating a strong host-versus-sandbox distinction at the individual-artifact level. Because these artifacts span multiple independent AI-environment layers, defenders must reproduce a realistic AI software ecosystem rather than merely scrub isolated indicators.
}

\subsection{RQ3: Causal Validation via Controlled Experiment}
\label{sec:controlled_experiment}
RQ2 identifies the AI-environment artifacts that discriminate between real hosts and sandbox environments; RQ3 investigates whether those artifacts arise causally from AI tools and package installation rather than incidental system configuration. To isolate causality, we perform a controlled cumulative installation experiment on a separate clean Windows~11 machine, progressively constructing a local AI ecosystem while repeatedly measuring the resulting artifact surface. The experiment proceeds in two phases. \textbf{Phase~A} installs a curated set of anchor AI tools one at a time to establish direct causal linkage between specific software packages and the resulting artifact observations. \textbf{Phase~B} performs large-scale automated package installation to evaluate how the measurable AI-environment surface expands as local tools and package dependencies accumulate. Across all installation steps, we execute the full probe suite together with the traditional VM-detection baseline, which serves as a negative control throughout the experiment.

\begin{table}[h]
\centering
\caption{Phase~A cumulative installation sequence. The installed tools collectively span multiple layers of the local AI dependency stack.}
\label{tab:phase_a_justification}
\small
\begin{tabular}{@{}llr@{}}
\toprule
\textbf{Step} & \textbf{Tool Installed} \\
\midrule
T01 & Ollama (port 11434, \texttt{.ollama/}) \\
T02 & Cursor IDE (\texttt{.cursor/}) \\
T02b & Claude Code (\texttt{.claude/}) \\
T02c & Gemini CLI (\texttt{.gemini/}) \\
T03 & HuggingFace (\texttt{.cache/huggingface/}) \\
T04 & PyTorch (\texttt{torch}) \\
T05 & LangChain + OpenAI + Anthropic SDKs \\
     & + \texttt{OPENAI\_API\_KEY} env var \\
T06 & Jupyter (\texttt{.jupyter/}, port 8888)\\
T07 & VSCode + GitHub Copilot (\texttt{.copilot/}) \\
\bottomrule
\end{tabular}
\end{table}

\subsubsection{Phase~A Anchor AI Tools Installation}
Figure~\ref{fig:growth_curve} and Table~\ref{tab:phase_a_results} in Appendix \ref{sec:appendix_rq2_phase_A_details} show that \tool detection rises monotonically as the local AI dependencies accumulate. Starting from a clean baseline, the system initially exhibits detection behavior comparable to sandbox environments, with no Class~C artifacts present: Topics-253 passes 23~artifacts (0.6\%), Trending-41 passes 3~artifacts (0.7\%), and SOTA detects 14~VM indicators (5.1\%). The baseline Topics-253 rate (0.6\%) differs slightly from the reference workstation's 0.41\% (Table~\ref{tab:ai_era_combined}) because the two environments have different Windows configurations; the discrepancy arises from generic Class~B artifacts rather than AI-specific tooling, as both machines contain zero Class~C artifacts at baseline. Each successive installation step expands the measurable AI-environment surface, increasing both project-level and individual-artifact detection rates, while the traditional VM-detection baseline remains largely unchanged throughout the experiment.


\begin{figure}[t]
\centering
\includegraphics[width=\columnwidth]{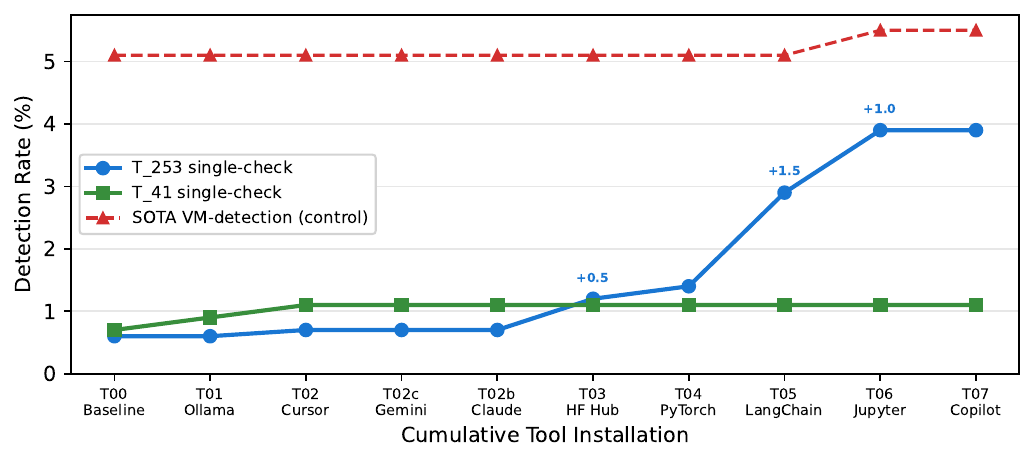}
\caption{Phase~A: \tool detection vs.\ anchor tool installation (10~steps). Topics-253 (solid) rises from 0.6\% to 3.9\%. SOTA (dashed) stays flat at ${\sim}$5.1\%. The full 214-step trajectory is shown in Figure~\ref{fig:full_growth_curve}.}
\label{fig:growth_curve}
\end{figure}

\PP{Dependencies Expansion Effects}
Lightweight AI-assistant tools contribute relatively few new artifacts: Ollama~(T01) adds +2 and Cursor~(T02) adds +3. Claude Code~(T02b) and Gemini CLI~(T02c) each produce $\Delta$=0 because \texttt{.claude/} and \texttt{.gemini/} appear in the Trending-41 corpus but not in the earlier Topics-253 crawl; both remain confirmed Class~C artifacts on the reference host (\S\ref{sec:eval_rq2}). In contrast, Python-based AI packages produce the largest expansion of the measurable artifact surface. LangChain/OpenAI SDK installation~(T05) contributes +59 artifacts through transitive dependencies (\eg \texttt{pydantic}, \texttt{tiktoken}, \texttt{aiohttp}), while Jupyter~(T06) adds +42 and HuggingFace~(T03) adds +21. These results suggest that application and package dependencies contribute more strongly to detectable AI-environment artifacts than the primary executable alone.

\PP{Why Trending-41 Single-Check Detection Plateaus}
The Trending-41 single-check rate plateaus at 1.1\% after T02, whereas Topics-253 continues rising from 0.7\% to 3.9\%. This difference reflects corpus composition rather than installation saturation. The curated Trending-41 dataset contains only 450 checks and is dominated by directories, ports, and environment variables, whereas the larger Topics-253 corpus (4{,}150 checks) contains substantially more Python-package artifacts. Consequently, package-heavy installations from T03 onward primarily expand the Topics-253 detection surface.

\PP{Traditional VM Detection Remains Largely Stable}
The traditional VM-detection baseline remains nearly unchanged throughout the experiment, holding at 14/273 indicators (5.1\%) through T01--T05 and increasing only slightly to 15/273 (5.5\%) after Jupyter installation (T06). The additional indicator likely arises from Jupyter's local network listener triggering a heuristic associated with VM-related services. Although minor (+0.4\%), this effect suggests that AI package installation can indirectly influence traditional anti-analysis heuristics.

\subsubsection{Phase~B: Extended Package Installation}
\begin{figure}[t]
\centering
\includegraphics[width=\columnwidth]{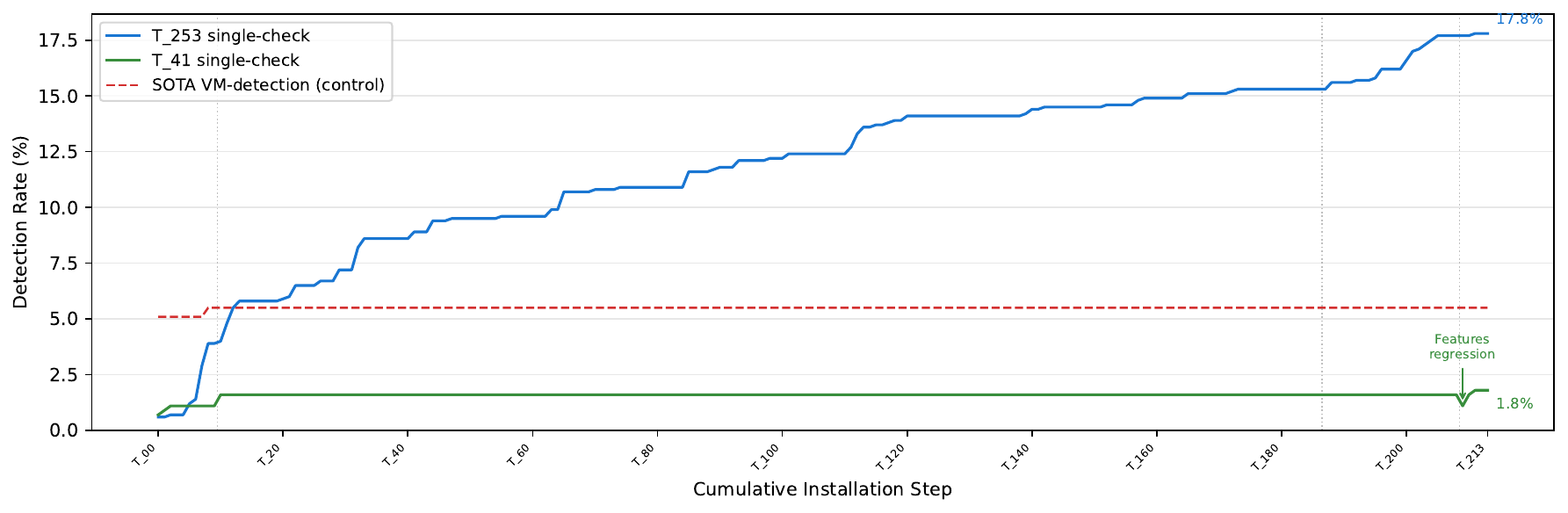}
\caption{Full growth trajectory across 214~installation steps.} 
\label{fig:full_growth_curve}
\end{figure}

Phase~A established direct causal linkage for 10 anchor AI tools. Phase~B extends this process through large-scale package accumulation: 199 automated package-installation attempts via pip, followed by 5 manual installations requiring Docker or specialized tooling. Figure~\ref{fig:full_growth_curve} shows the complete 214-step trajectory. Topics-253 single-check detection rises from 3.9\% at the end of Phase~A to 15.3\% after bulk installation and reaches 17.8\% following retries and manual installs---a nearly 30$\times$ increase over the 0.6\% clean baseline. The resulting growth curve exhibits a staircase pattern: successful installations introduce bursts of transitive dependencies, followed by temporary plateaus until additional packages are installed. Of the 199 pip-install attempts, 100 succeed (50.3\%); many failures arise because GitHub repository names do not always map directly to installable pip package names. The final artifact distribution shows that \textbf{Python packages dominate the measurable AI-environment surface}. Of 735 passing artifact checks
(17.8\% of the 4{,}126 installable artifacts derived from the original 4{,}150 crawled artifacts), 719 are Python packages, compared to only 7 directories, 4 environment variables, and 5 open ports. This imbalance arises because individual package installations recursively introduce large transitive dependencies that appear across many projects' artifact sets. At the project level, the final installation state detects 184/253 Topics projects (72.7\%) and 30/41 Trending projects (73.2\%). In contrast, the traditional VM-detection baseline remains effectively unchanged throughout all 214 installation steps, fluctuating only between 5.1\% and 5.5\%. This persistent stability reinforces that traditional VM indicators and AI-environment artifacts capture fundamentally different environmental properties.

\summarybox{
\Tool detection is causally tied to AI tools and package installation: a clean machine begins at sandbox-equivalent levels (0.6\%) and rises monotonically to 17.8\% across 214 installation steps, while the traditional VM-detection baseline remains nearly unchanged (${\sim}$5.1\%). Many detectable artifacts emerge from transitive dependencies introduced by the primary tools.
}

\subsection{RQ4: Defense Evaluation via Artifact Spoofing}
\label{sec:eval_rq4}
In RQ4, we evaluate whether analysts can realistically reduce the observed host-versus-sandbox asymmetry through artifact spoofing. We study two questions: (1) whether progressively richer spoofing strategies can mitigate AI environment detection, and (2) whether an adaptive attacker can distinguish spoofed environments from genuine AI-capable environments.

\PP{Spoofing Setup}
Using the same clean Windows~11 baseline and probe infrastructure from RQ3, we evaluate two cumulative spoofing strategies with progressively richer spoofing behavior (Table~\ref{tab:defense_levels}). \textbf{Class-C (minimal spoofing):} spoofs the 8 most commonly referenced Class-C artifacts---6~directories (\texttt{.cursor}, \texttt{.claude}, \texttt{.copilot}, \texttt{.gemini}, \texttt{.cache/huggingface}, \texttt{.ollama}), 1~environment variable (\texttt{OPENAI\_API\_KEY}), and 1~port (11434). We excluded four less consistently discriminative Class-C artifacts (\texttt{.conda}, \texttt{.jupyter}, \texttt{CUDA\_PATH} directory, port~8888) from the minimal strategy because they have lower cross-project reference frequency and are more likely to appear on non-AI hosts. \textbf{Full-Coverage Spoofing:} spoofs every per-project check instance that failed the baseline scan---69~directory checks (from 19~unique paths), 1{,}362~environment variable checks (from 144~unique variables), 143~port checks (from 28~unique ports), and 46~Python package checks. The counts exceed unique artifact counts because the same artifact (e.g., \texttt{OPENAI\_API\_KEY}) appears in multiple projects' check lists; spoofing one unique variable satisfies all project-level references. After each spoofing level, we execute the full probe suite and compare spoofed environments against both the clean baseline and the genuine AI-tool installation state constructed in RQ3.

\begin{table}[h]
\centering
\caption{Cumulative defense levels for artifact spoofing.}
\label{tab:defense_levels}
\small
\setlength{\tabcolsep}{4pt}
\begin{tabular}{@{}clll@{}}
\toprule
\textbf{Lv.} & \textbf{Action} & \textbf{Cost} & \textbf{Example} \\
\midrule
L1 & Create empty dirs & Trivial & \texttt{.claude/}, \texttt{.ollama/} \\
L2 & + Set env vars & Trivial & \texttt{OPENAI\_API\_KEY=sk-...} \\
L3 & + Open ports & Low & Ports 11434, 8888 \\
L4 & + Install packages & Moderate & \texttt{torch}, \texttt{langchain} \\
L5 & + Populate data & High & Model files, config JSONs \\
\bottomrule
\end{tabular}
\end{table}

\begin{figure}[h]
\centering
\includegraphics[width=\columnwidth]{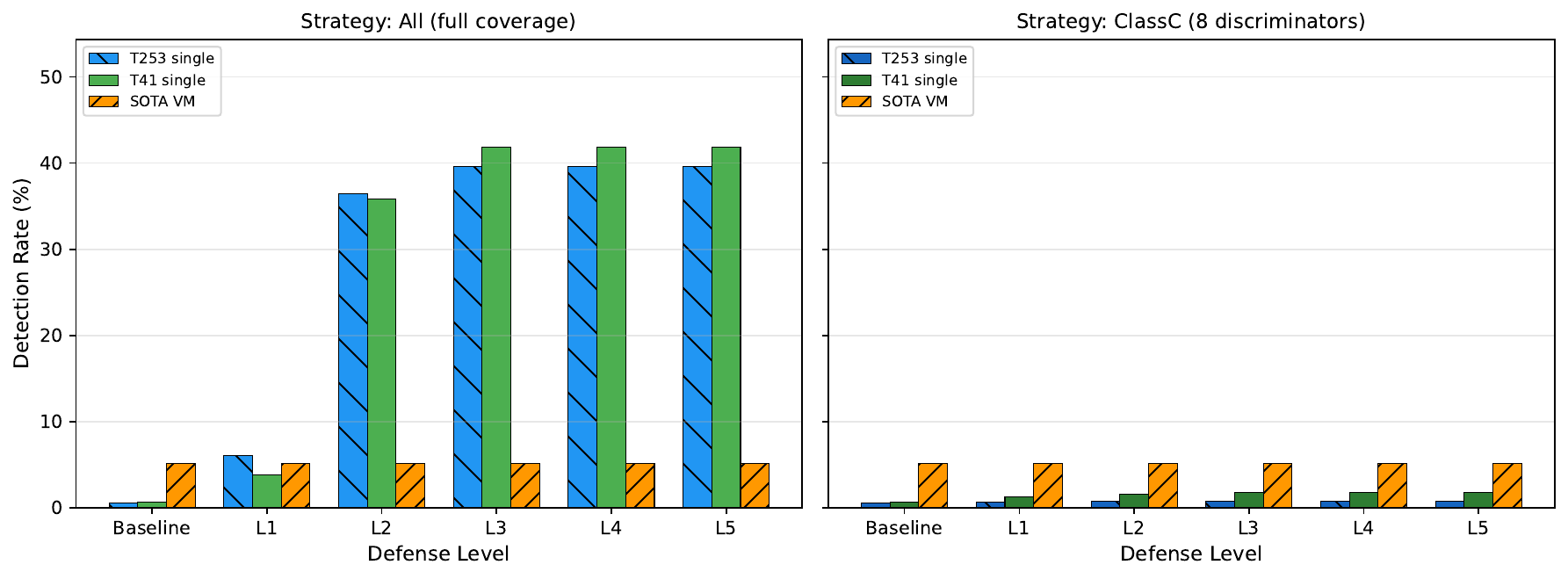}
\caption{Detection under progressive spoofing: full-coverage defenses plateau at L3, lightweight Class-C spoofing remains near baseline, and traditional VM-detection rates stay unchanged throughout.}
\label{fig:defense_comparison}
\end{figure}

\begin{table}[h]
\centering
\caption{Probe detection under All defense (every failing artifact spoofed). Detection plateaus at L3.}
\label{tab:defense_all}
\small
\begin{tabular}{@{}lrrrrc@{}}
\toprule
 & \multicolumn{2}{c}{\textbf{T253}} & \multicolumn{2}{c}{\textbf{T41}} & \textbf{SOTA} \\
\cmidrule(lr){2-3} \cmidrule(lr){4-5}
\textbf{Level} & \textbf{Single} & \textbf{Proj} & \textbf{Single} & \textbf{Proj} & \textbf{VM} \\
\midrule
Baseline & 0.6\% & 82 & 0.7\% & 12 & 5.1\% \\
L1 (dirs) & 6.1\% & 132 & 3.8\% & 19 & 5.1\% \\
L2 (+env) & 36.4\% & 202 & 35.8\% & 28 & 5.1\% \\
L3 (+port) & 39.6\% & 231 & 41.8\% & 33 & 5.1\% \\
L4--L5 & 39.6\% & 231 & 41.8\% & 33 & 5.1\% \\
\bottomrule
\end{tabular}
\end{table}

\subsubsection{Defender Results}
Tables~\ref{tab:defense_all} quantify detection behavior under progressively richer spoofing strategies. Shallow spoofing improves project-level detection substantially but exhibits diminishing returns. Under the All strategy, environment-variable spoofing (L2) produces the largest increase, raising Topics-253 detection from 6.1\% to 36.4\% because many project-level checks reference shared variables. In contrast, lightweight Class-C spoofing as shown in Table~\ref{tab:defense_classc} in Appendix~\ref{sec:rq4spoof} yields only modest gains because it introduces comparatively few AI-environment artifacts. As shown in Figure~\ref{fig:defense_comparison}, both strategies reach ${\sim}$40\% at L3 and subsequently plateau.
Attempts to extend spoofing through user-space Python-package installation provide no measurable additional benefit because package placement and runtime structure remain inconsistent with genuine AI-tool installations. These results suggest that reproducing realistic AI-environment structure requires more than creating superficial artifacts: it demands coherent AI software ecosystems with consistent dependency structure, runtime behavior, and filesystem layout.

\subsubsection{Adaptive Attacker Response}
We next evaluate whether an adaptive attacker can distinguish spoofed artifacts from genuine installations through lightweight structural validation. For each spoofing level, we implement consistency checks that inspect expected filesystem structure, credential formatting, protocol behavior, package coherence, and cache realism. Across all defense levels, structural validation correctly identifies 4 of 5 tested artifacts as spoofed or incomplete as shown in Appendix~\ref{ap:attacker} Table~\ref{tab:structural_verdicts}. The only exception, \texttt{.claude/}, remained an installation artifact from the earlier RQ3 environment and therefore represents a conservative evaluation artifact rather than a spoofing success. Collectively, these results indicate that shallow artifact spoofing can inflate detection metrics, but reproducing structurally consistent AI ecosystems remains operationally expensive.

\subsubsection{Operational Cost Asymmetry}
Beyond detection effectiveness, spoofing introduces an asymmetric operational burden. Table~\ref{tab:cost_asymmetry} compares the defender effort required to reproduce realistic AI-environment artifacts against the lightweight structural checks required for attacker-side validation. Effort Level in Table~\ref{tab:cost_asymmetry} approximates the lines of code required for lightweight structural checks that distinguish spoofed artifacts from genuine installations. For example, at L1, an attacker only needs $\sim$5 lines of code to detect a non-empty directory, which can be executed in under 3 seconds. Other levels may require slightly more code and more involved checks (e.g., artifacts like models or JSON files in L5), but all remain substantially cheaper than reproducing the corresponding AI-environment artifacts. These results expose a structural asymmetry in fourth-generation sandbox evasion: attackers validate realism through lightweight consistency checks, while defenders must reproduce and maintain increasingly coherent AI software ecosystems. The cost of simulation, therefore, grows substantially faster than the cost of detection.

\begin{table}[h]
\centering
\caption{Cost asymmetry for Class-C defense.}
\label{tab:cost_asymmetry}
\small
\begin{tabular}{@{}lrrrl@{}}
\toprule
\textbf{Level} & \textbf{Setup} & \textbf{Effort Level} & \textbf{Artifact Check} \\
 & \textbf{(sec)} & \textbf{(LoC)} & \\
\midrule
L1 & $<$3 & 5 & Dir non-empty \\
L2 & $<$2 & 3 & Key prefix format \\
L3 & $<$6 & 8 & HTTP req + JSON \\
L4 & 0 & 0 & Import + version \\
L5 & $\sim$180 & 15 & Timestamps, content \\
\bottomrule
\end{tabular}
\end{table}

\summarybox{
Artifact spoofing exhibits a fundamental cost asymmetry: validating AI-environment realism requires only lightweight structural checks, whereas reproducing convincing AI ecosystems demands coherent runtimes, dependencies, services, and ongoing ecosystem maintenance. As a result, reproducing realistic AI environments becomes operationally more expensive than detecting shallow spoofing.
}

\section{Discussion}
\label{sec:discussion}
Our evaluation results for RQ1--RQ4 suggest that AI-capable software ecosystems introduce a new realism gap for sandboxed execution environments. 

\PP{Defense Implications}
RQ4 in particular exposes a fundamental operational asymmetry: validating AI-environment realism requires only lightweight structural checks, whereas reproducing convincing AI ecosystems demands increasingly complex and continuously maintained software stacks. Even aggressive spoofing strategies remain distinguishable from genuine installations due to inconsistencies across artifacts, runtimes, and dependency state. This burden is amplified in heterogeneous multi-backend infrastructures such as VirusTotal, where inconsistent provisioning across backends itself becomes a fingerprinting signal. These findings motivate future research on AI-aware sandbox environments that more closely reflect real-world AI ecosystems.

\PP{Stealth of Probing}
The underlying checks rely exclusively on common Win32 APIs such as \texttt{PathFileExists()}, \texttt{GetEnvironmentVariable()}, and \texttt{connect()}, all of which are routinely invoked by legitimate applications during startup. Consequently, detecting \tool-style probing behaviorally may prove difficult without introducing substantial false positives, though we leave empirical measurement of such rates to future work.

\PP{Limitations}
\label{sec:limitations}
Our study has several limitations. First, the proposed AI-based evasion is inherently targeted: non-AI-capable hosts may remain indistinguishable from sandboxes under these checks alone. Second, our evaluation compares sandbox backends against a limited number of reference systems. Third, the extracted artifact set represents a snapshot of the rapidly evolving AI ecosystem as of February~2026. Finally, while recent malware already probes GPUs and AI APIs~\cite{google2025honestcue, lummastealer2025, desai2016antivmsandbox}, we do not claim confirmed in-the-wild adoption of the specific \tool probes studied here. Future work includes large-scale prevalence measurement across real endpoints, cross-platform evaluation on Linux and macOS, longitudinal tracking of sandbox adaptation, and hybrid classifiers that combine \tool artifacts with traditional wear-and-tear features.

\section{Conclusion}
\label{sec:conclusion}

We presented the first systematic measurement study of AI-environment artifacts as a new sandbox-evasion surface. Through artifact extraction from 284 GitHub AI projects, cross-sandbox evaluation across seven backends, controlled causal validation, and adaptive spoofing analysis, our results show that \tools discriminate where traditional VM-detection baselines fail. Moreover, reproducing convincing AI ecosystems proves substantially more expensive than detecting shallow spoofing, creating a fundamental operational asymmetry for sandbox operators. More broadly, our findings suggest that sandbox realism is increasingly shaped not only by hardware fidelity or simulated user activity, but also by the software ecosystems installed on the host. As AI-capable environments continue to proliferate across developer and enterprise systems, maintaining realistic AI-environment fidelity may become an increasingly important challenge for future malware-analysis infrastructure.






\balance
\bibliographystyle{IEEEtran}
\bibliography{ref}


\appendices

\section{Sandbox Evasion Generations}\label{ap:gen}
\begin{table}[H]
 \centering
 \caption{Sandbox evasion generations. Each exploits a different realism gap between genuine endpoints and sandbox environments, requiring progressively more operationally expensive defenses (maintenance burden shown in parentheses).}
 \label{tab:generation_comparison}
 \small
 \setlength{\tabcolsep}{5pt}
 \begin{tabularx}{\columnwidth}{@{}>{\raggedright\arraybackslash}X>{\raggedright\arraybackslash}X>{\raggedright\arraybackslash}X@{}}
 \toprule
 \textbf{Generation} & \textbf{Artifact type} & \textbf{Defense (cost)} \\
 \midrule
 1st: VM artifacts & Registry, CPUID, MAC & Scrub indicators (one-time) \\
 2nd: Timing \& emulation & Clock skew, TSC, cache behavior & Bare-metal / high-fidelity execution (one-time) \\
 3rd: Wear-and-tear realism & Browser history, event logs \& user activity & Synthetic aging (periodic) \\
 \textbf{4th: \Tools AI-environment realism (ours)} & \textbf{Dirs, env vars, runtimes, ports, packages, model caches} & \textbf{Install \& maintain AI ecosystem (ongoing} \\
 \bottomrule
 \end{tabularx}
 \end{table}
\section{Clustering Algorithm}
\label{ap:appendix_algorithm}

\begin{algorithm}[H]
\caption{Artifact Clustering by Project}
\label{alg:clustering}
\begin{algorithmic}[1]
\Require $\mathcal{P}$: set of projects with artifacts $P.A$
\Ensure Per-project confidence scores
\For{\textbf{each} project $P \in \mathcal{P}$}
    \State $n \gets 0$
    \For{\textbf{each} artifact $a \in P.A$}
        \If{$a.type = \textsc{Dir}$}
            \State $n \gets n + \Call{FileExists}{a}$
        \ElsIf{$a.type = \textsc{Env}$}
            \State $n \gets n + \Call{GetEnv}{a}$
        \ElsIf{$a.type = \textsc{Port}$}
            \State $n \gets n + \Call{TcpConnect}{a}$
        \ElsIf{$a.type = \textsc{Pkg}$}
            \State $n \gets n + \Call{FileExists}{a}$
        \EndIf
    \EndFor
    \State $conf \gets n \;/\; |P.A|$
    \State \Call{Output}{$P$, $conf$}
\EndFor
\end{algorithmic}
\end{algorithm}

\section{DNS Protocol Details}
\label{ap:dns}

This appendix provides detailed technical documentation of our DNS-based exfiltration protocol. The protocol overview is presented in Section~\ref{sec:dns_exfil}.

\subsection{Protocol Format}

Each exfiltration session begins and ends with marker queries:

\smallskip
\noindent\footnotesize
\texttt{Start: s\{sid\}-start-\{chunks\}-\{flag\}.\{domain\}}\\
\texttt{End:\;\; s\{sid\}-end-\{chunks\}.\{domain\}}
\normalsize
\smallskip\\

Each data chunk uses the format:

\smallskip
\noindent\footnotesize
\texttt{\{hex1\}.\{hex2\}.s\{session\}-c\{chunk\}.\{domain\}}
\normalsize
\smallskip

\noindent where \texttt{hex1} and \texttt{hex2} are two 60-character hex labels (30~bytes each). Each DNS query carries \textbf{60~bytes} of binary data. Chunks are sent to multiple Interactsh domains (default:~3) for redundancy.

\subsection{LZNT1 Compression}
\label{sec:lznt1}

Payloads are compressed using Windows' native LZNT1 algorithm via \texttt{ntdll.dll}'s \texttt{RtlCompressBuffer()}, achieving 60--80\% size reduction. Compressed payloads use an 8-byte header: 4-byte magic (\texttt{ZLIB}), 4-byte original size.

\begin{table}[h]
\centering
\caption{Compression ratios for typical payloads.}
\small
\begin{tabular}{@{}lrrr@{}}
\toprule
\textbf{Report Type} & \textbf{Original} & \textbf{Compressed} & \textbf{Reduction} \\
\midrule
SOTA (269 checks) & 27 KB & 8 KB & 70\% \\
Topics 253 single & 85 KB & 25 KB & 71\% \\
Trending 41 project & 12 KB & 4 KB & 67\% \\
\bottomrule
\end{tabular}
\end{table}

\subsection{Timing and Reliability}
The protocol uses 200ms inter-chunk delay, 2 retries per chunk per domain, and 3$\times$ redundancy on markers. For a 150-chunk report: $\text{Time} \approx 150 \times 200\text{ms} \times 2 = 60$~seconds. Multi-domain merging recovers data when individual domains lose chunks.

\begin{table}[H]
\centering
\caption{DNS exfiltration protocol parameters.}
\small
\begin{tabular}{@{}ll@{}}
\toprule
\textbf{Parameter} & \textbf{Value} \\
\midrule
Bytes per query & 60 (two 30-byte hex labels) \\
Compression & LZNT1 (Windows native, 60--80\% reduction) \\
Domains & 3 (redundancy) \\
Retries per chunk & 2 per domain \\
Inter-chunk delay & 200ms \\
Collection service & Interactsh (open-source) \\
\bottomrule
\end{tabular}
\end{table}

DNS exfiltration successfully retrieved results from all seven sandbox platforms (ANY.RUN, CAPEv2, Cuckoo, Hybrid Analysis, Kaspersky, Triage, VirusTotal) with 100\% data recovery.

\section{RQ2 Additional Experiments}
\label{sec:appendix_rq2_details}

\subsection{Threshold Sensitivity}

Table~\ref{tab:threshold_sensitivity} confirms that the project-level discrimination gap is robust across detection thresholds. Even at the strictest threshold (50\%), the reference host detects 10 projects vs.\ at most 5 for any sandbox backend.

\begin{table}[H]
\centering
\caption{Threshold sensitivity for project detection (Trending~41).}
\label{tab:threshold_sensitivity}
\small
\begin{tabular}{@{}lcccc@{}}
\toprule
\textbf{Bar} & \textbf{Real Host} & \textbf{CAPEv2} & \textbf{Best (vt\_v4)} & \textbf{Min Gap} \\
\midrule
5\% & 25/41 (61.0\%) & 5/41 (12.2\%) & 14/41 (34.1\%) & +26.9\% \\
\textbf{15\%} & \textbf{19/41 (46.3\%)} & \textbf{3/41 (7.3\%)} & \textbf{11/41 (26.8\%)} & \textbf{+19.5\%} \\
25\% & 16/41 (39.0\%) & 2/41 (4.9\%) & 8/41 (19.5\%) & +19.5\% \\
50\% & 10/41 (24.4\%) & 1/41 (2.4\%) & 5/41 (12.2\%) & +12.2\% \\
\bottomrule
\end{tabular}
\end{table}

\subsection{Phase~A cumulative installation Details}
\label{sec:appendix_rq2_phase_A_details}

\begin{table}[H]
\centering
\caption{Phase~A cumulative installation results. Detection rises monotonically. ``$\Delta$'' counts artifacts that flipped from \textsc{Fail} to \textsc{Pass} at each step.}
\label{tab:phase_a_results}
\small
\begin{tabular}{@{}llrrrr@{}}
\toprule
\textbf{Step} & \textbf{Tool} & \textbf{T253} & \textbf{T41} & \textbf{$\Delta$} & \textbf{SOTA} \\
\midrule
T00 & Baseline           & 0.6\% & 0.7\% & ---  & 5.1\% \\
T01 & Ollama             & 0.6\% & 0.9\% & +2   & 5.1\% \\
T02 & Cursor             & 0.7\% & 1.1\% & +3   & 5.1\% \\
T02b & Claude Code       & 0.7\% & 1.1\% & ---  & 5.1\% \\
T02c & Gemini CLI        & 0.7\% & 1.1\% & ---  & 5.1\% \\
T03 & HuggingFace        & 1.2\% & 1.1\% & +21  & 5.1\% \\
T04 & PyTorch            & 1.4\% & 1.1\% & +11  & 5.1\% \\
T05 & LangChain+OpenAI   & 2.9\% & 1.1\% & +59  & 5.1\% \\
T06 & Jupyter            & 3.9\% & 1.1\% & +42  & 5.5\% \\
T07 & VSCode+Copilot     & 3.9\% & 1.1\% & +1   & 5.5\% \\
\bottomrule
\end{tabular}
\end{table}

\subsection{Project-Level Clustering}
Project-level clustering reproduces the same asymmetry observed at the artifact level. Core AI ecosystems such as CUDA, Claude Code, and Ollama exhibit substantially higher confidence scores on the reference workstation than on sandbox backends, while detections on sandbox platforms are primarily driven by generic Class artifacts rather than AI-specific tooling. Across all evaluated thresholds, the reference workstation consistently detects substantially more AI-project ecosystems than contemporary sandbox environments.

\begin{table}[h]
\centering
\caption{Per-project detection: reference workstation vs.\ CAPEv2. Core AI tools show high confidence on the host and zero on the sandbox.}
\label{tab:project_detection}
\small
\begin{tabular}{@{}lcc@{}}
\toprule
\textbf{Project} & \textbf{Real Host} & \textbf{CAPEv2} \\
\midrule
nvidia/cuda-toolkit & 60.0\% & 0\% \\
anthropics/claude-code & 50.0\% & 0\% \\
microsoft/vscode-copilot & 50.0\% & 0\% \\
ollama/ollama & 42.9\% & 0\% \\
anaconda/anaconda & 28.6\% & 0\% \\
huggingface/transformers & 20.0\% & 0\% \\
openai/openai-python & 20.0\% & 0\% \\
langchain-ai/langchain & 12.5\% & 0\% \\
\midrule
\textbf{Total Detected ($\geq$15\%)} & \textbf{19/41} & \textbf{3/41} \\
\bottomrule
\end{tabular}
\end{table}

\subsection{VirusTotal Backend Heterogeneity}
VirusTotal's multi-backend architecture reveals substantial internal variation in project-level detection behavior. Across the same probe binary, backend detection rates range from 0\% to 26.8\% (Table~\ref{tab:vt_heterogeneity}), indicating that individual backend environments expose different subsets of generic AI-environment artifacts. However, despite this variability, no VirusTotal backend contains any Class~C artifact. Thus, while project-level clustering varies across backend instances, host-exclusive AI-environment artifacts maintain consistent sandbox-versus-host discrimination across the entire VirusTotal infrastructure.

\begin{table}[h]
\centering
\caption{VirusTotal backend heterogeneity across identical probe submissions. Project-level detection rates vary substantially across backend instances, while Class~C host-exclusive artifacts remain absent from all backends. ``---'' denotes data not collected.}
\label{tab:vt_heterogeneity}
\small
\setlength{\tabcolsep}{3pt}
\begin{tabular}{@{}lcccc@{}}
\toprule
\textbf{Backend} & \textbf{41 Proj} & \textbf{253 Proj} & \textbf{450 Single} & \textbf{41 Single} \\
\midrule
vt\_v0 & 7.3\% & 11.5\% & 0.2\% & 0.07\% \\
vt\_v1 & 0.0\% & 11.5\% & 0.2\% & 0.07\% \\
vt\_v2 & 7.3\% & 2.4\% & 1.3\% & 0.36\% \\
vt\_v3 & 7.3\% & 0.4\% & 0.2\% & 0.05\% \\
vt\_v4 & 26.8\% & 11.9\% & 0.0\% & 0.07\% \\
vt\_v5 & 0.0\% & --- & --- & 0.29\% \\
vt\_v6 & 0.0\% & --- & --- & --- \\
\bottomrule
\end{tabular}
\end{table}

\section{RQ4 Additional Results}
\subsection{RQ4 Probe Detection Under Class-C and All Defense}\label{sec:rq4spoof}
\begin{table}[H]
\centering
\caption{Probe detection under Class-C defense (8 discriminating artifacts only).}
\label{tab:defense_classc}
\small
\begin{tabular}{@{}lrrrrc@{}}
\toprule
 & \multicolumn{2}{c}{\textbf{T253}} & \multicolumn{2}{c}{\textbf{T41}} & \textbf{SOTA} \\
\cmidrule(lr){2-3} \cmidrule(lr){4-5}
\textbf{Level} & \textbf{Single} & \textbf{Proj} & \textbf{Single} & \textbf{Proj} & \textbf{VM} \\
\midrule
Baseline & 0.6\% & 82 & 0.7\% & 12 & 5.1\% \\
L1 & 0.7\% & 83 & 1.3\% & 14 & 5.1\% \\
L2 & 0.8\% & 113 & 1.6\% & 18 & 5.1\% \\
L3 & 0.8\% & 117 & 1.8\% & 20 & 5.1\% \\
L4--L5 & 0.8\% & 117 & 1.8\% & 20 & 5.1\% \\
\bottomrule
\end{tabular}
\end{table}

\subsection{Structural Attacker Check Details}\label{ap:attacker}
Table~\ref{tab:structural_verdicts} provides detailed structural check verdicts across defense levels. Each row represents a Class-C artifact; each column shows the structural check outcome at that defense level. \texttt{.claude} passes at all levels because it is a genuine installation from the RQ3 experiment. All spoofed artifacts are correctly identified at every level.

\begin{table}[h]
\centering
\caption{Structural check verdicts (All strategy). \cmark{} = genuine, all others correctly flagged as spoofed/absent.}
\label{tab:structural_verdicts}
\footnotesize
\setlength{\tabcolsep}{3pt}
\begin{tabular}{@{}lccccc@{}}
\toprule
\textbf{Artifact} & \textbf{L1} & \textbf{L2} & \textbf{L3} & \textbf{L4} & \textbf{L5} \\
\midrule
\texttt{.claude}   & \cmark & \cmark & \cmark & \cmark & \cmark \\
\texttt{OPENAI\_..} & ---    & Fmt    & Fmt    & Fmt    & Fmt \\
Port 11434          & ---    & ---    & NoHTTP & NoHTTP & NoHTTP \\
\texttt{torch}      & ImpF   & ImpF   & ImpF   & Abs    & Abs \\
\texttt{.cache/hf}  & NoMdl  & NoMdl  & NoMdl  & NoMdl  & Size \\
\bottomrule
\end{tabular}
\\[2pt]
{\scriptsize \cmark\,=\,genuine OK;\; Fmt\,=\,bad format;\; NoHTTP\,=\,no HTTP response;\; ImpF\,=\,import fail;\; Abs\,=\,absent;\; NoMdl\,=\,no models;\; Size\,=\,too small;\; ---\,=\,not applicable.}
\end{table}


\end{document}